\documentclass[aps, prx, twocolumn, showpacs, superscriptaddress]{revtex4-1}
\usepackage{graphicx}
\usepackage{amsmath}
\usepackage{amssymb}
\usepackage{physics}
\usepackage{hyperref}
\usepackage{float}
\usepackage{lipsum}
\usepackage{ulem}
\hypersetup{colorlinks=true,
	    final=true,
	    linkcolor=blue,
	    citecolor=blue,
	    filecolor=blue,
	    urlcolor=blue,
}

\begin{document}
\title{Conductivity of infinite-layer NdNiO$_{2}$ as a probe of spectator bands}
\author{Harrison LaBollita}
\email{hlabolli@asu.edu}
\affiliation{Department of Physics, Arizona State University, Tempe, AZ 85287, USA}
\author{Alexander Hampel}
\affiliation{Center for Computational Quantum Physics, Flatiron Institute, 162 5th Avenue, New York, NY 10010, USA}
\author{Jonathan Karp}
\affiliation{Department of Applied Physics and Applied Math, Columbia University, New York, NY, 10027, USA}
\author{Antia S. Botana}
\affiliation{Department of Physics, Arizona State University, Tempe, AZ 85287, USA}
\author{Andrew J. Millis}
\affiliation{Center for Computational Quantum Physics, Flatiron Institute, 162 5th Avenue, New York, NY 10010, USA}
\affiliation{Department of Physics, Columbia University, New York, NY, 10027, USA}
\date{\today}

\begin{abstract}
Using a density-functional theory plus dynamical mean-field theory methodology, we compute the many-body electronic structure and optical conductivity of NdNiO$_{2}$ under the influence of large scattering rates on the Nd($5d$) bands and including dynamical interactions on the Nd($5d$) orbitals with shifts of the Nd-Ni $d$-level energy difference. We find a robust conducting pathway in the out-of-plane direction arising from strong hybridization between the Ni-$d_{z^2}$ and Nd($5d$) orbitals. This pathway can be ``short-circuited'' if this hybridization is suppressed through large electronic scattering rates but is not reduced to zero even by very large beyond-DFT shifts of the Nd-Ni $d$-level energy splitting. The computed in-plane conductivity for NdNiO$_{2}$ predicts the material to be a ``good metal'' in  contrast to experiments indicating the material is a ``bad metal'' or ``weak insulator''. Our results motivate future experiments measuring the $c$-axis resistivity as a proxy for the spectator bands and suggests the essential difference between the infinite-layer nickelates and the cuprates is dimensionality of their electronic structures.
\end{abstract}
\maketitle

\section{\label{sec:intro}Introduction}

The discovery of superconductivity in infinite-layer \cite{Li2019superconductivity, Osada2020superconducting, Zeng2021superconductivity, Osada2021nickelate} and quintuple-layer \cite{Pan2021superconductivity} nickelates establishes the ``$d^9$'' layered nickel compounds with the generic chemical formula $R_{n+1}$Ni$_{n}$O$_{2n+2}$ ($n \geq 2$; $R$ = La, Pr, Nd) as a novel family of materials that can provide new insights into superconductivity. In particular, the structural and chemical similarities to the high-T$_{c}$ cuprates \cite{Bednorz1986possible} suggest that the materials may provide an important perspective on electronically mediated high transition temperature superconductivity.

Despite the structural and electronic similarities between the cuprates and nickelates, there are many important differences in their physical properties \cite{Botana2020similarities,Lechermann2020late, Lechermann2020multiorbital, Lechermann2021doping,Karp2020manybody, Chen2022Elect, Liu2020electronic, Choi2020role, Bandyopadhyay2020superconductivity,Kang2021optical, Kapeghian2020electronic, Krishna2020effects, Been2021electronic, Chen2022dynamical}. The parent-phase of the cuprates is an antiferromagnetic insulator with a $\sim$ 1.5 eV charge gap and a room temperature resistivity $\gtrsim 100$ m$\Omega$cm, while  the stoichiometric infinite-layer nickelates are at most only weakly insulating with $\rho \sim 1$ m$\Omega$cm near room temperature \cite{Li2020superconducting, Osada2020phase, Lee2022character} and no reported evidence of long-range antiferromagnetic (AFM) order \cite{Fowlie2022Intri, Hayward1999sodium}.

\begin{figure}
    \centering
    \includegraphics[width=\columnwidth]{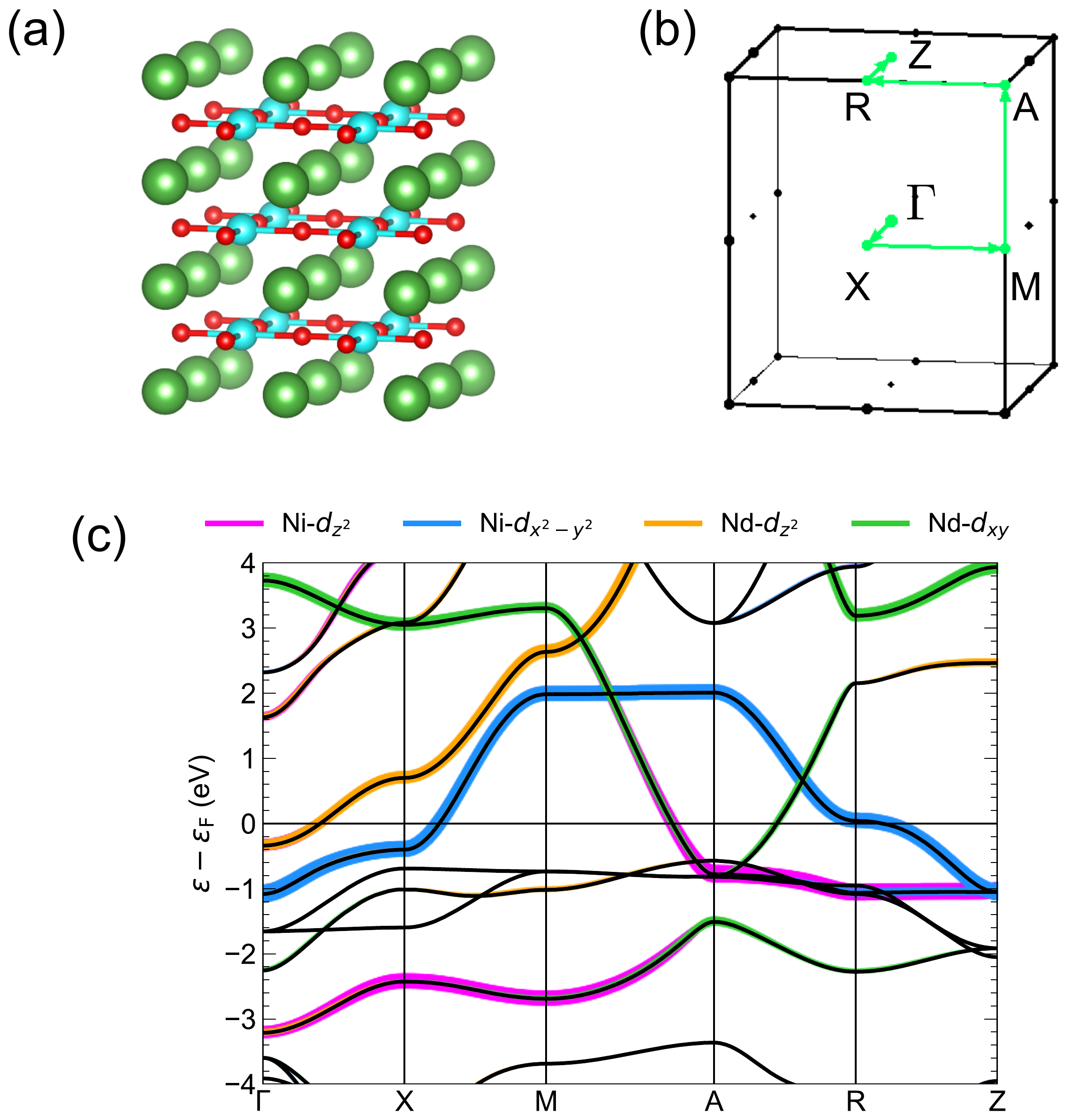}
    \caption{(a) Crystal structure of infinite-layer NdNiO$_{2}$ with Nd (green), Ni (cyan), and O (red) atoms. (b) Brillouin zone with the path along high-symmetry directions shown in green for plots of the DFT band structure $\varepsilon_{\nu}({\bf k})$ and spectral function $A({\bf k}, \omega)$. (c) Band structure of NdNiO$_{2}$ within DFT with orbital character shown for Nd-$d_{z^{2}}$, Nd-$d_{xy}$, Ni-$d_{z^{2}}$, and Ni-$d_{x^{2}-y^{2}}$ depicted.}
    \label{fig:structure}
\end{figure}

Basic quantum chemical (formal valence) arguments as well as density functional theory (DFT) electronic structure calculations indicate that in both the high-T$_c$ cuprates and the layered nickelates the transition metal ion (Cu or Ni) is in or near a $d^9$ valence state with the hole in the $d$-shell residing in the $d_{x^2-y^2}$ orbital so that an important feature of the DFT-level electronic structure of both the cuprate and nickelate materials is a quasi-two-dimensional transition metal derived $d_{x^{2}-y^{2}}$ band crossing the Fermi level \cite{Keimer2015From, Pickett1989electronic}. In the cuprate materials the $d_{x^{2}-y^{2}}$-derived band is the only relevant near Fermi surface band, whereas the electronic structure of the nickelates includes additional bands Nd($5d$) orbitals hybridized with other Ni $d$-orbitals (see Fig. \ref{fig:structure}) \cite{Pickett2004infinite, Botana2020similarities}. These additional bands, sometimes referred to as ``spectator'' or ``self-doping" bands, are a crucial difference between the cuprate and nickelate materials, and their role in the low-energy physics of the infinite-layer nickelate is a subject of great current interest \cite{Lechermann2020multiorbital, Botana2020similarities, Karp2020comparative, Karp2020manybody, Goodge2020doping, Petocchi2020normal, Louie2022twogap}. At minimum the spectator bands affect the low energy physics  by changing the relation between the carrier density in the Ni $d_{x^2-y^2}$  band and the chemical composition, so that the stoichiometric NdNiO$_2$ compound has a fractionally filled $d_{x^2-y^2}$ band and is not Mott insulating. However, it is possible that the ``spectator'' bands play a more important role in the physics, for example by allowing other $d$-orbital character near the Fermi level, opening the possibility of ``Hund's physics'' \cite{Kang2021optical}.

In this paper,  we present the results of computational experiments designed to shed light on the physics of the spectator bands and on one of the observables that may enable experimental determination of their role. We show that the different components of the conductivity tensor are sensitive reporters of the presence and physics of spectator bands, and therefore present the conductivities following from each many-body electronic structure calculations in comparison to the experimental resistivity data \cite{Li2020superconducting}. Our calculations study two possibilities. First, we consider what may be termed the ``standard model" of the nickelate materials in which  all of the interesting physics is carried by the Ni($3d$) orbitals (specifically, the $x^{2}-y^{2}$ orbital), which are treated as correlated (within DMFT) and give rise to physics rather similar to that of the cuprates, while the other bands are treated on a non-interacting (DFT) level.  In these ``standard model" computations we add in addition a phenomenological scattering rate to the ``spectator band" states; tuning this rate to a large value effectively removes the contributions of the spectator bands to the transport enabling a determination of transport signatures of the spectator bands.  In a second set of computational experiments we include interactions both on  the Ni and on the Nd $d$ states and manipulate the many-body electronic structure  via an adjustment of double counting potentials using a charge self-consistent combination of density-functional theory and dynamical mean-field theory (DFT+DMFT) in order to determine whether the presence of the spectator bands is a robust consequence of the level on which the beyond-DFT correlations are treated. 

The rest of this paper is organized as follows. In Section~\ref{sec:methods} we present the  theoretical and computational methods. In Section~\ref{sec:results} we present the ``standard model'' electronic structure and optical conductivity of NdNiO$_{2}$ assuming many-body correlations only on the Ni orbitals but controlling the contribution of the Nd orbitals to transport via a phenomenological scattering rate (Sec. \ref{sec:exp1}) and then we consider a more general interacting model with correlations included also on the Nd($5d$) orbitals within DMFT (Sec. \ref{sec:exp2}).  Section~\ref{sec:summary} is a summary and conclusion, indicating also directions for future research.

\section{\label{sec:methods}Methodology}

We use different charge self-consistent density-functional theory plus dynamical mean-field theory (DFT+DMFT) frameworks to compute the many-body electronic structure and optical conductivity of NdNiO$_{2}$. The computations use the experimental lattice parameters of the tetragonal P4/mmm symmetry  stoichiometric  NdNiO$_{2}$ compound \cite{Li2019superconductivity} (see Fig. \ref{fig:structure}a).

For DFT we used the all-electron, full potential augmented plane-wave plus local orbital (APW+lo) basis set method as implemented in {\sc wien}2k \cite{Blaha2020wien2k} with the Perdew-Burke-Ernzerhof (PBE) version \cite{Perdew1996generalized} of the generalized gradient approximation (GGA) for the exchange-correlation functional. A dense $k$-mesh of 40$\times$40$\times$40 is used for integration in the Brillouin zone for the self-consistent calculations. We used $R_{\mathrm{MT}}K_{\mathrm{max}} = 7$ and muffin-tin radii of 2.5, 1.93, and 1.71 a.u. for Nd, Ni, and O, respectively. The Nd($4f$) states are treated as core states. 

For the DMFT calculations, we construct either a single impurity problem for the full 5-orbital Ni($3d$) shell or two impurity problems: one for the 5-orbital Ni($3d$) shell and one for the 2-orbitals Nd($5d$) orbitals \{Nd-$d_{z^{2}}$, Nd-$d_{xy}$\} which participate in the fermiology of this material. Both scenarios are treated within the single-site DMFT approximation.  The atomic-like orbitals are created via the projection method \cite{Aichhorn2016dfttools, Aichhorn2009interface} with a large energy window of size $-10$ to $10$ eV around the Fermi level. 

We determine the interaction parameters by calculating the static Coulomb interaction $U(\omega=0)$ within the constrained random phase approximation (cRPA)~\cite{Aryasetiawan2004} as implemented in \textsc{VASP}~\cite{Kresse:1993bz,Kresse:1996kl,Kresse:1999dk}. The Coulomb matrix elements are evaluated from maximally localized Wannier functions (MLWF)~\cite{PhysRevB.80.155134} using \textsc{Wannier90}~\cite{Mostofi_et_al:2014}. To obtain similarly localized orbitals as used in DMFT we construct Wannier functions in a large energy window for all Ni($3d$), Nd($5d$), and O($2p$) orbitals. To evaluate the constrained polarization function we use the projection scheme via the constructed MLWFs~\cite{Kaltak2015}. The cRPA calculation is performed on a $9 \times 9 \times 9$ ${\bf k}$-mesh (plus finite size corrections), with $\sim$360 empty bands, and using a plane wave cut off of 333~eV when evaluating the polarization function. The resulting Coulomb tensor is then symmetrized in the Ni($3d$) and Nd($5d$) sub-block to obtain the interaction parameters. The interactions on the Ni impurity are governed by the rotationally-invariant Slater Hamiltonian parameterized by the Hubbard $U = F^{0} = 7.1$ eV and Hund's coupling $J_{\mathrm{H}} =\frac{1}{14}(F^{2}+F^{4}) = 1$ eV. For the Nd impurity, we apply an appropriate two orbital Hubbard-Kanamori Hamiltonian with $U = 4.2$ eV and $J_{\mathrm{H}} = 0.44$ eV.  The fully-localized limit (FLL) formula is used for the double counting correction, which has the following form: $\Sigma_{\mathrm{DC}} = \frac{1}{2}U^{\prime}N(N-1)$ with $N$ being the total occupation of the Ni or Nd site. The term proportional to $J$ is not written. Throughout this work, we lave used $U^{\prime} = U$ unless otherwise indicated.

\begin{figure*}
\centering
\includegraphics[width=2\columnwidth]{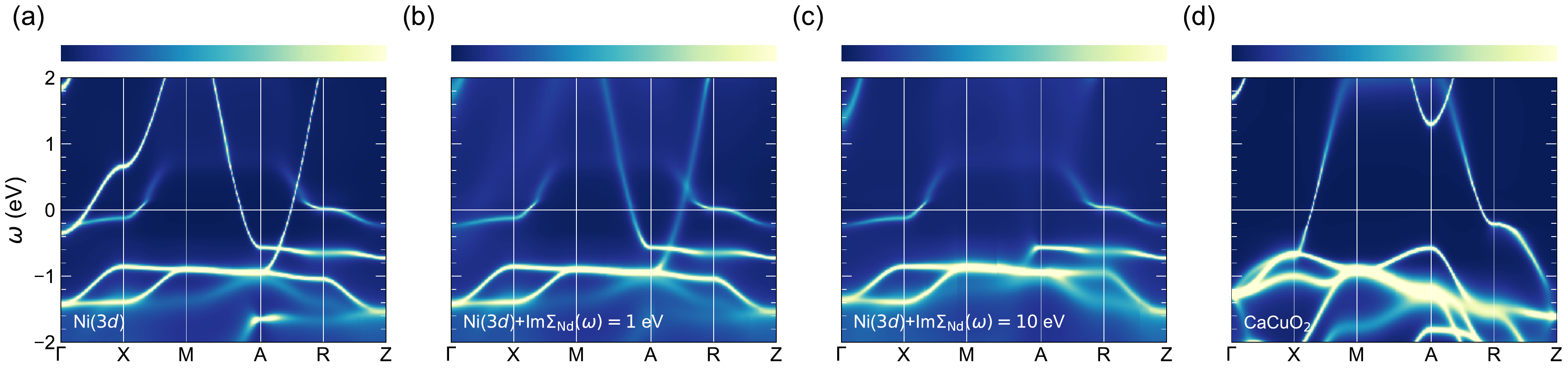}
\caption{Summary of momentum resolved spectral data, $A({\bf k}, \omega)$, along high-symmetry lines in the Brillouin zone for a DFT+DMFT calculation with the 5 orbital Ni($3d$) shells treated as correlated with constant self-energies applied to the Nd($5d$) states. (a) $\Sigma_{\mathrm{Nd}}(\omega) = 0$, (b) $\Sigma_{\mathrm{Nd}}(\omega) = -i$ eV, and (c) $\Sigma_{\mathrm{Nd}}(\omega) = -10i$ eV. (d) Spectral function $A({\bf k}, \omega)$ for CaCuO$_{2}$ in the metallic state.}
\label{fig:spectral-const-sig}
\end{figure*}

We employ a continuous-time quantum Monte Carlo (QMC) algorithm based on the hybridization expansion method as implemented in TRIQS/{\sc cthyb} \cite{Parcollet2015triqs, Priyanka2016cthyb} to solve the Ni and Nd impurity problems. To reduce high-frequency noise in the QMC data, we represent the Green's function in a basis of Legendre polynomials and sample the Legendre coefficients directly within the TRIQS/{\sc cthyb} solver \cite{Boehnke2011orthogonal}. All calculations are performed at a system temperature of 290 K ($\beta = 40$ eV$^{-1}$) in the paramagnetic state. Maximum entropy methods are used to analytically continue the QMC data and the diagonal components of the self-energy from Matsubara space to real-frequency space \cite{Kraberger2017maxent}.

From the electronic structure obtained from the various DFT+DMFT calculations, we compute the frequency dependent optical conductivity within the Kubo formalism, as implemented in the TRIQS/DFTtools software package \cite{Aichhorn2016dfttools}. The locality of the DMFT self-energy means that vertex corrections may be neglected. The frequency dependent optical conductivity is given by 

\begin{eqnarray}
\sigma^{\alpha\beta}(\Omega) = N_{\mathrm{sp}}\pi e^{2}\hbar&& \int d\omega \, \Gamma_{\alpha\beta}(\omega+\Omega/2, \omega-\Omega/2)\\
&&\times ~~ \frac{f(\omega-\Omega/2) - f(\omega+\Omega/2)}{\Omega}
\nonumber
\end{eqnarray}
where
\begin{equation}
\Gamma_{\alpha\beta}(\omega, \omega') = \frac{1}{V} \sum_{{\bf k}} \mathrm{Tr} \Big [v^{\alpha}({\bf k})A({\bf k},\omega)v^{\beta}({\bf k})A({\bf k}, \omega') \Big ]
\end{equation}
The spectral function $A$ and velocity operator $v$ are tensors in the space of band indices and the velocity operator in direction $\alpha \in \{ x, y, z\}$ is
\begin{equation}
v^{\alpha}_{\nu\nu'}({\bf k}) = -i \langle \psi_{\nu}({\bf k}) | \nabla^{\alpha} | \psi_{\nu'}({\bf k}) \rangle/m_{e}
\end{equation}
the matrix elements of $v$ are computed within the WIEN2k optics code \cite{Sofo2006w2koptics} on a dense $60 \times 60 \times 60$ ${\bf k}$-mesh. For numerical stability, we use a broadening of 10 meV for the calculation of all optical conductivity data.

\section{Results \label{sec:results}}

\subsection{\label{sec:exp1}Basic electronic structure and scattering effects}
Fig \ref{fig:structure}c shows the non-interacting (DFT) band structure of NdNiO$_{2}$ calculated within DFT for NdNiO$_{2}$ along the high-symmetry path in the Brillouin zone shown in Fig~\ref{fig:structure}b. The orbital characters of the bands are highlighted. The basic features of the low-energy physics of NdNiO$_{2}$, as described in previous works, are revealed \cite{Botana2020similarities,Karp2020manybody,Pickett2004infinite}. A quasi-two-dimensional Ni-$d_{x^{2}-y^{2}}$ derived band crosses Fermi level and is analogous to the Cu-$d_{x^{2}-y^{2}}$ band found in DFT calculations of the cuprates. In addition, the DFT calculation reveals an electron pocket of mainly Nd-$d_{z^{2}}$ character centered at the $\Gamma$ point and a second pocket, of mixed Ni-$d_{z^2}/Nd_{xy}$ character centered at the A point \cite{Botana2020similarities,Labollita2021electronic}. These two bands accept carriers from the Ni-$d_{x^2-y^2}$ band; the consequences of this ``self-doping" effect are still up for debate \cite{Lechermann2020multiorbital, Botana2020similarities, Karp2020comparative, Karp2020manybody, Goodge2020doping, Petocchi2020normal}. 

\begin{figure}
    \centering
    \includegraphics[width=\columnwidth]{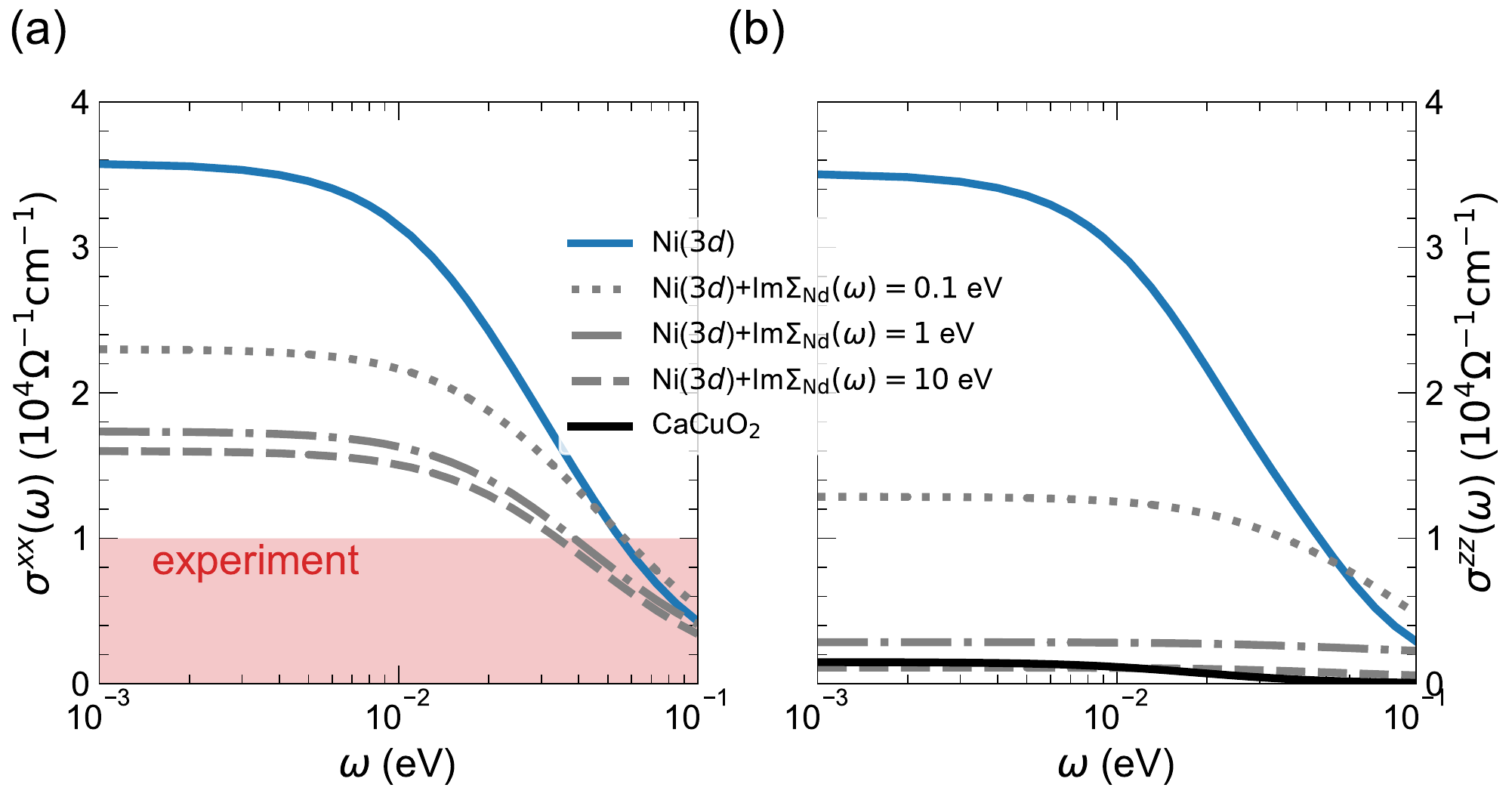}
    \caption{Calculated optical conductivity from DFT+DMFT (Ni($3d$) correlated) (a) in-plane and (b) out-of-plane with different scattering rates applied to the Nd($5d$) orbitals. Additionally, optical conductivity for CaCuO$_{2}$ in the out-of-plane is given for comparison in (b). The experimental DC conductivity is denoted by the shaded red region. }
    \label{fig:optical-const-sig}
\end{figure}

\begin{figure*}
\centering
\includegraphics[width=1.8\columnwidth]{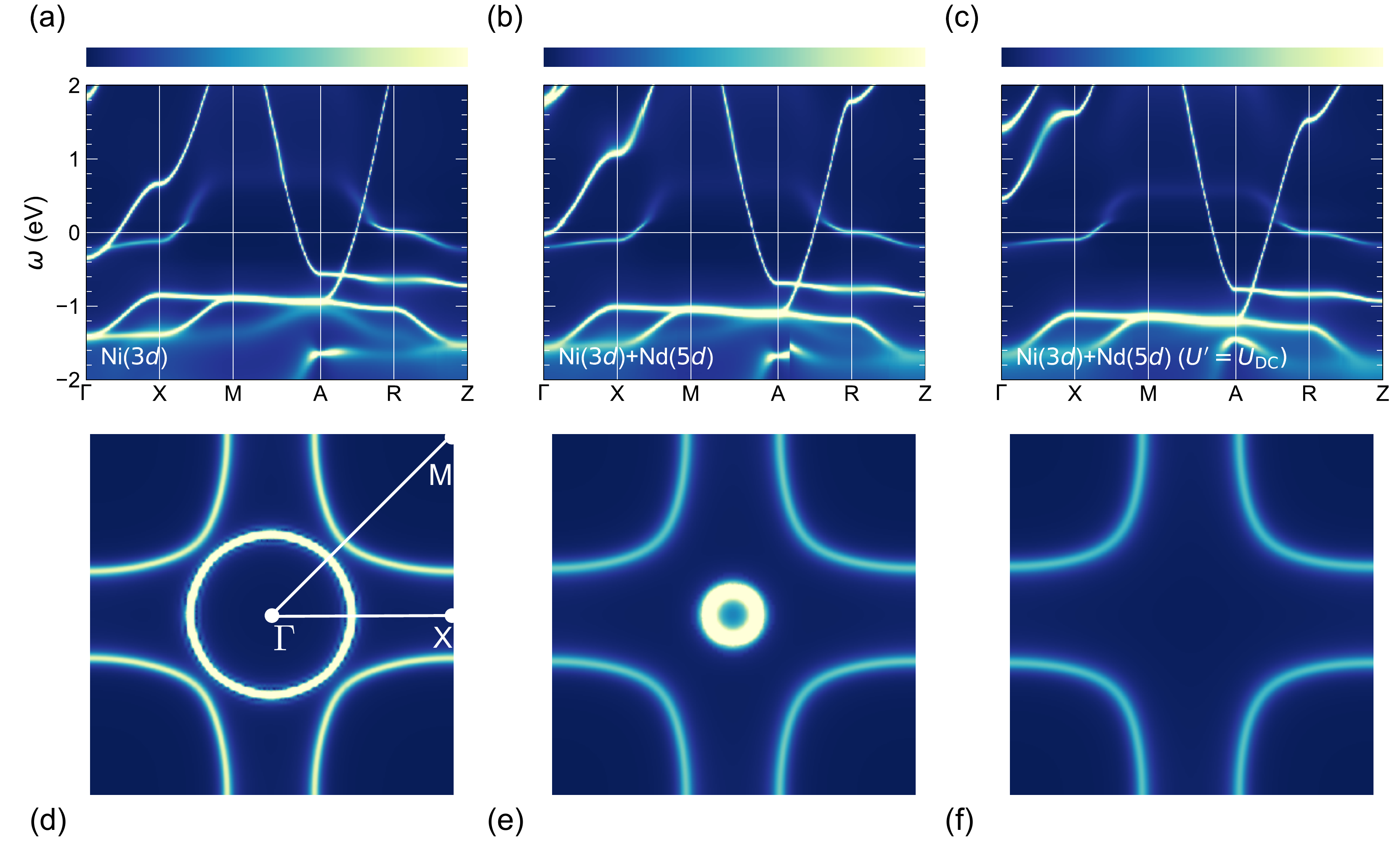}
\caption{Summary of momentum resolved spectral data, $A({\bf k}, \omega)$, along high-symmetry lines in the Brillouin zone. (a) Conventional DFT+DMFT calculation with the 5 orbital Ni($3d$) shells treated as correlated. (b) DFT+DMFT calculation where the 5 orbital Ni($3d$) shell and the Nd-$\{d_{z^{2}}, d_{xy} \}$ orbitals treated as correlated. (c) Same as (b) with an adjusted double counting term $U_{\mathrm{DC}} = U^{\prime}$. (d-f) Corresponding interacting Fermi surfaces for (a-c), respectively, in the $k_{z} = 0$ plane where the A-centered pocket does not cross the chemical potential.}
\label{fig:spectral-dc}
\end{figure*}

Using a DFT+DMFT framework where the five Ni($3d$) orbitals are treated as correlated and all others are treated as uncorrelated, we perform many-body electronic structure calculations for NdNiO$_{2}$. We then introduce an additional electronic scattering onto the Nd($5d$) states by adding a constant self-energy of the form $\Sigma_{\mathrm{Nd}}(\omega) = -i\eta$, where $\eta$ is a real, positive constant.

Figure \ref{fig:spectral-const-sig} presents the results as a false-color plot of the trace of the band-basis many-body spectral function for three choices of phenomenological scattering rate. Fig. \ref{fig:spectral-const-sig}a presents the standard model results with no broadening of the Nd-derived bands. The band structure is similar to the DFT band structure, except that the Ni-$d_{x^2-y^2}$ band is substantially narrowed. Panel Fig.~\ref{fig:spectral-const-sig}d presents the results of an analogous DFT+DMFT calculation for CaCuO$_{2}$ which is isostructural and chemically analogous to NdNiO$_{2}$. Within the single-site DMFT approximation the accepted interaction parameters ($U=7$ eV, $J_{\mathrm{H}}=1$ eV) leave CaCuO$_2$ in its metallic state and antiferromagentism or cluster DMFT methods are needed to reproduce the observed insulating behavior \cite{Karp2020manybody, Karp2022super}.  Importantly, there is no hybridization between the Cu($3d$)-Ca($3d$) orbitals exhibited by no dispersing band present between the M and A points (see Fig.~\ref{fig:spectral-const-sig}d). For $\Sigma_{\mathrm{Nd}} = -1i$ eV (Fig. \ref{fig:spectral-const-sig}b), the Nd-$d_{z^2}$ band becomes so strongly broadened that the $\Gamma$-centered spectator pocket is no longer visible while the A spectator pocket, although broadened, remains visible, in part because of the admixture of Ni $d_{z^2}$ states. Figure \ref{fig:spectral-const-sig}c shows that for $\Sigma_{\mathrm{Nd}} = -10i$ eV all of the Nd($5d$) states are so broadened that only the single Ni-$d_{x^{2}-y^{2}}$ band is visible at the chemical potential.

Figure \ref{fig:optical-const-sig} summarizes the  optical conductivity in the low-frequency regime computed from the different spectral functions shown in Fig. \ref{fig:spectral-const-sig}. For the in-plane conductivity (Fig. \ref{fig:optical-const-sig}a), we find that additional scattering effects on the Nd($5d$) states decreases the Drude peak with a maximum decrease by about 50\% for $\Sigma_{\mathrm{Nd}} = -10i$ eV showing that the spectator bands contribute about half of the in-plane conductivity in the standard model. The out-of-plane conductivity shows a similar systematic trend with a much stronger depletion dropping to essentially zero at larger scattering rates showing that the spectator bands completely control the out of plane conductivity.

This conclusion is reinforced by the CaCuO$_2$ out of plane conductivity also shown in Fig.~\ref{fig:optical-const-sig}b, which is  nearly zero for CaCuO$_{2}$ and matches the calculation for NdNiO$_{2}$ when the hybridizing $k_{z}$ band is destroyed via a large scattering rate. The in-plane conductivity for CaCuO$_{2}$ (not shown) is significantly larger than NdNiO$_{2}$, which is a consequence of the different Fermi velocities at the chemical potential from the $d_{x^{2}-y^{2}}$-derived bands in each material (compare Fig.~\ref{fig:spectral-const-sig}a to Fig.~\ref{fig:spectral-const-sig}d). This highlights an important electronic difference between the cuprate and infinite-layer nickelate in terms of the strong $c$-axis coupling exhibited by the nickelate.

\begin{figure}
\centering
\includegraphics[width=\columnwidth]{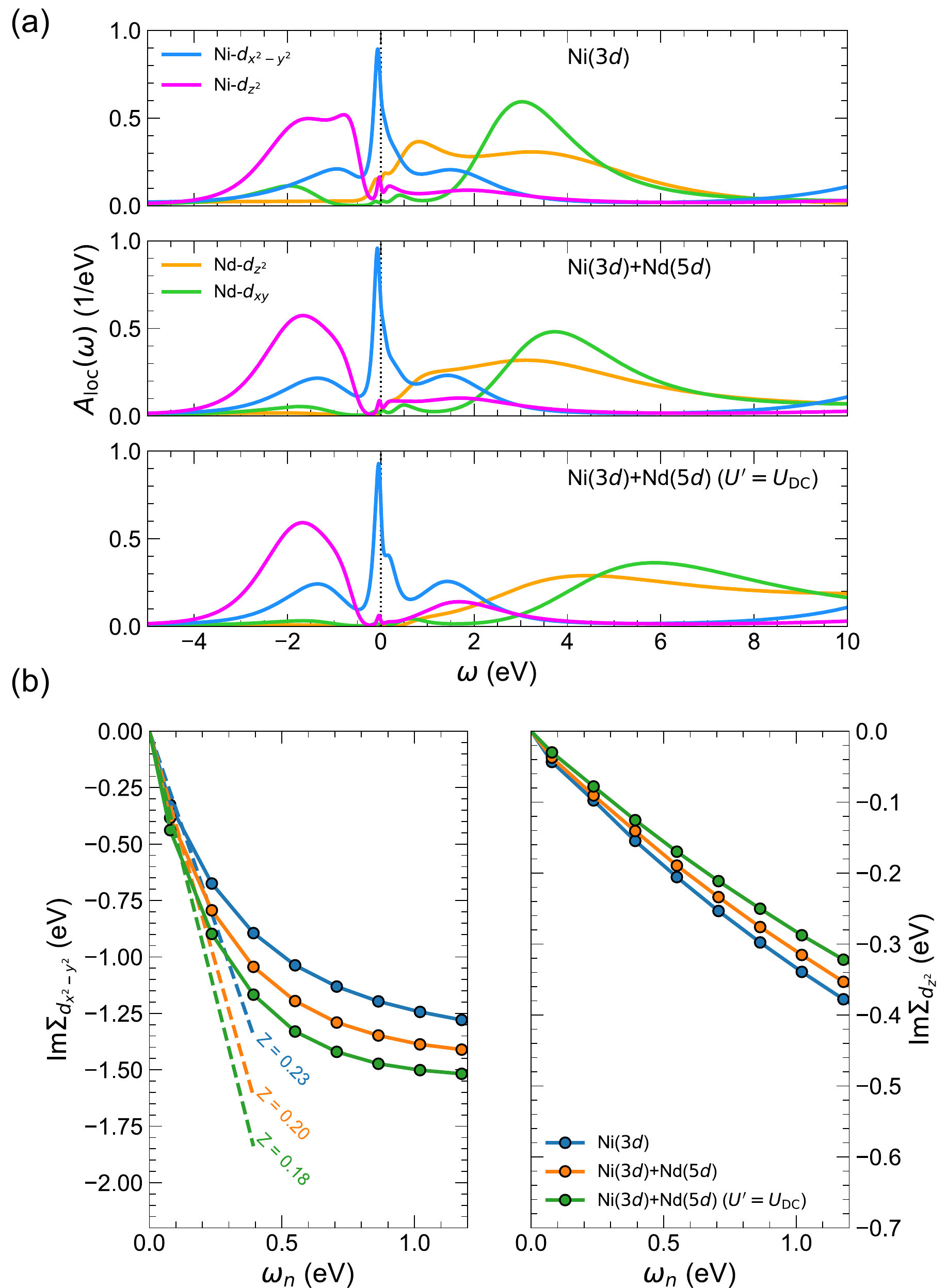}
\caption{(a) Local spectral functions for the Ni (only $e_{g}$ orbitals shown) and Nd impurity problems for three different computational experiments performed. (b) Matsubara self-energies for the Ni-$e_{g}$ orbitals: $d_{x^{2}-y^{2}}$ (left) and $d_{z^{2}}$ (right).}
\label{fig:local-spectral}
\end{figure}

\begin{figure}
    \centering
    \includegraphics[width=\columnwidth]{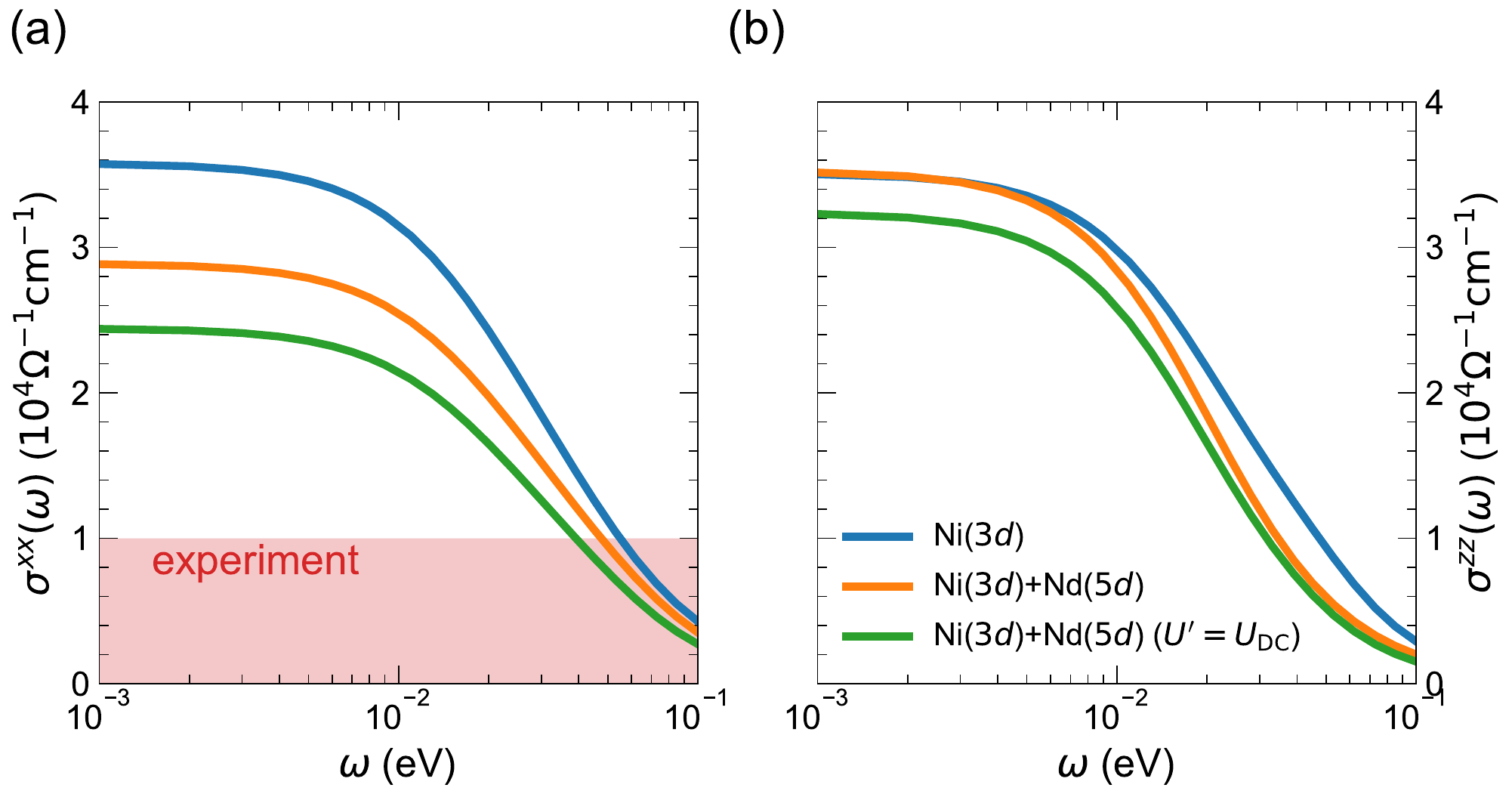}
    \caption{Calculated optical conductivity for different DFT+DMFT electronic structure theories (a) in-plane and (b) out-of-plane. The experimental DC conductivity is denoted by the shaded red region.}
    \label{fig:optical-dcU}
\end{figure}

\subsection{\label{sec:exp2}Dynamical interactions applied to the Nd($5d$) orbitals}
This subsection presents the results of a second computational experiment, where we add electronic correlations within DMFT to the Nd($5d$) orbitals ($d_{z^{2}}$, $d_{xy}$), with interaction parameters obtained from cRPA, to understand how additional beyond-DFT interactions might influence the electronic structure of NdNiO$_{2}$ and how these effects might be revealed in the optical conductivity. Figure \ref{fig:spectral-dc} summarizes the ${\bf k}$-resolved spectral data for the second set of computational experiments. Figure \ref{fig:spectral-dc}a is repeated from Fig. \ref{fig:spectral-const-sig}a for reference. The corresponding Fermi surface in the $k_{z} = 0$ plane is shown in Fig. \ref{fig:spectral-dc}d. 

Figure \ref{fig:spectral-dc}b shows the spectral function, $A({\bf k}, \omega)$ from a DFT+DMFT calculation with the 5 Ni($3d$) and 2 Nd($5d$) orbitals treated as correlated (referred to as Ni($3d$)+Nd($5d$)). Interestingly, the $\Gamma$-pocket band arising from the Nd-$d_{z^{2}}$ has been shifted up with respect to the overall spectrum. The strongly $k_{z}$ dispersing hybridized Nd band centered at the $A$-point remains essentially unchanged between the two calculations (Fig. \ref{fig:spectral-dc}a vs. Fig. \ref{fig:spectral-dc}b). The interacting Fermi surface reveals the significant reduction in the size of the Nd-$d_{z^{2}}$ electron pocket (see Fig. \ref{fig:spectral-dc}e). The area of the Ni-$d_{x^{2}-y^{2}}$ sheet remains essentially constant.

In DFT+DMFT calculations the double counting term plays a crucial role especially for transition-metal oxides \cite{Park2014computing, Wang2012covalency}. Operationally, this term controls the relative energy shift of the correlated states with respect to the uncorrelated states with the goal of canceling out the correlation contributions encoded in the DFT exchange-correlation functional. In Fig. \ref{fig:spectral-dc}c, we alter the double counting potential to attempt to displace the Nd($5d$) states away from the chemical potential by setting $U^{\prime} = U_{\mathrm{DC}} = 10$ eV $\sim 2U_{\mathrm{Nd}}$. This changes the double counting potential $\Sigma_{\mathrm{DC}}$ from $-0.78$ eV for $U^{\prime} = U_{\mathrm{Nd}}$ to $-3.1$ eV for $U^{\prime} = U_{\mathrm{DC}}$. This treatment results in a complete removal of the Nd-$d_{z^{2}}$ electron pocket. However, the hybridizing $k_{z}$ band remains unchanged. We note that this is in contrast to the mechanism of a constant scattering rate shown in Fig.~\ref{fig:spectral-const-sig}c. The Fermi surface sheets generated from the Ni-$d_{x^{2}-y^{2}}$ bands remain constant across all calculations (see Fig. \ref{fig:spectral-dc}(d-f)).

Figure~\ref{fig:local-spectral} further reveals the low-energy spectrum for these three different calculations. Across the three calculations, the quasiparticle spectral weight of the Ni-$d_{x^{2}-y^{2}}$ states remains dominant around the chemical potential. The Nd-$d_{z^{2}}$ states gradually decrease and eventually become fully depleted at $\omega = 0$ when the double counting potential is adjusted. Spectral weight from the hybridized Ni-$d_{z^{2}}$ states also decreases at the chemical potential in connection to the Nd-$d_{z^{2}}$ removal. The electronic correlations captured quantitatively from the quasiparticle renormalization factor $Z = (1 - \partial\mathrm{Im}\Sigma/\partial \omega_{n} \big |_{\omega_{n}\rightarrow 0})$ increase for the Ni-$d_{x^{2}-y^{2}}$ as the Nd($5d$) states are pushed further away from the chemical potential (see Fig. \ref{fig:local-spectral}b). This reveals the proximity of the Ni-$d_{x^{2}-y^{2}}$ to a Mott-like state that may be masked by the presence of the hybridized Nd($5d$)-Ni-$d_{z^{2}}$ states \cite{Gu2020substantial}. 

We now turn to the resulting optical conductivity shown in Fig. \ref{fig:optical-dcU}. The Drude peak for the conductivity in the plane ($\sigma_{xx}$) shows a systematic decrease across the three different DFT+DMFT calculations for NdNiO$_{2}$ (see Fig. \ref{fig:optical-dcU}a). There are two contributions to this decrease. First, the the in-plane conductivity contributions from the Nd($5d$) states are systematically removed. Second, the electronic correlations of the Ni-$d_{x^{2}-y^{2}}$ increase resulting in a smaller (larger) quasiparticle renormalization $Z$ (mass enhancement $m^{\star}/m$). In the low-frequency regime, the optical conductivity reads \cite{Petocchi2020normal}:
\begin{equation}
\mathrm{Re}\,\sigma(\omega) = \frac{\sigma_{\mathrm{DC}}}{\pi} \frac{\tau}{1 - (\omega\tau)^{2}} + \sigma_{\mathrm{inc}}(\omega),
\end{equation}
where $\sigma_{\mathrm{DC}}  = (Z n )e^{2}$ and $n$ is the carrier density. Thus, lowering $Z$ decreases the Drude peak $\sigma_{\mathrm{DC}}$. Interestingly, the out-of-plane conductivity remains essentially unchanged across the three different calculations as shown in Fig. \ref{fig:optical-dcU}b. In all three of these calculations, there is no mechanism that alters this hybridizing $k_{z}$ band which offers the only route for conduction out-of-plane. Therefore, $\sigma_{zz}$ remains mostly unchanged.

\section{\label{sec:summary}Summary and Discussion}
One motivation for the current interest in NdNiO$_2$ and related materials is the perspective that these compounds provide on the relationship between superconductivity and electron correlation effects. The family of nickelate materials shares with the family of cuprate materials a low energy electronic structure prominently featuring a two dimensional approximately half-filled band derived from the transition metal $d_{x^2-y^2}$ orbital and subject to strong interactions. An important difference between the two material families is the presence, in the Ni-compounds, of ``spectator bands" derived from Nd $d$-orbitals that are present near the Fermi level and provide both a three dimensional dispersion (in the case of the infinite-layer nickelate) or strong interlayer coupling (in the case of the 3 and 5 layer compounds) and a ``self doping" effect in which the occupancy of the $x^2-y^2$ derived band becomes non-integer even at stoichiometric compositions and holes are introduced into the $d_{3z^2-r^2}$ orbital. There are two spectator bands: one centered at the $\Gamma$ point derived from  Nd $d_{3z^2-r^2}$ and apparently weakly coupled to the Ni states, and one centered at the A point derived from Nd $d_{xy}$ states and coupled to Ni $d_{3z^2-r^2}$ states.

Understanding the role of the ``spectator bands" in the physics of the materials is an important open question.  This paper explores the spectator band issue via a set of computational experiments that treat the correlations on the Nd site on the same level of theory as the correlations on the Ni site, consider various modifications of the standard theory that change the contributions of the spectator bands to the low energy physics, and present the frequency dependent conductivity (especially the interband conductivity) as an important experimental diagnostic of the effects of the spectator bands because the Nd orbitals provide a robust conducting pathway along out-of-plane direction (in the infinite layer material NdNiO$_2$) or low-lying interband transitions (in related materials such as $R_{n+1}$Ni$_{n}$O$_{2n+2}$ ($n \neq \infty$) where a $R$O$_{2}$ slab cuts the $c$-axis dispersion \cite{Labollita2021electronic, Labollita2022manybody}).

We find, consistent with previous results \cite{Kang2021optical, Petocchi2020normal}, that the standard correlation theory which treats the Nd orbitals as weakly correlated predicts that NdNiO$_2$ is a good metal with rather isotropic conductance, in contrast to experiment which shows that the in-plane conductance is large, characteristic of a ``bad metal" (the out of plane conductance is not known). Modifying the model by adding a large ad-hoc scattering rate to the Nd orbitals completely suppresses the interplane conductivity, but reduces the in-plane conductivity by only a factor of two or so without changing the theoretically predicted good metal behavior. 

We then extended the theory to treat correlations on the Nd sites on the same DFT+DMFT level as the correlations on the Ni sites. The interactions seem to deplete the $\Gamma$ pocket, leaving the minimum energy of this band very close to the Fermi level, while not significantly changing the A pocket. We further adjusted the relative energies of the Ni and Nd $d$-states by changing the ``double counting correction" in the calculation. With modest adjustment the $\Gamma$ pocket can be entirely removed \cite{Si2020topo} but the A pocket is robust even to large changes, so that  at this level of theory the strong $c$-axis coupling is not altered by electronic correlation effects. Indeed, previous studies have shown that in the vicinity of the A pocket there is significant hybridization also with interstitial states that do not have a clear atomic origin \cite{Gu2020substantial, Chen2022dynamical,Lechermann2020multiorbital}. These interstitial states would not be subject to local correlation effects, perhaps accounting for the resilience of the A pocket. Furthermore, our adjustment of the spectator band states acts as a ``governor'' on the correlations of the strongly correlated Ni-$d_{x^{2}-y^{2}}$ states, masking via the self-doping effect a potential nearby Mott-like state.

In summary, the computational experiments performed here show that both the ``standard model" DFT+DMFT approach (correlations only on the Ni site) and any reasonable deformation of it lead to a  calculated in-plane DC conductivity that is  incompatible with available experiments, because the spectator bands cannot be eliminated from the low energy theory. The dc and optical interplane conductivity, as well as angle-resolved photoemission experiment measurements especially of the A-pocket, are important tests of the theory. On the theoretical side, our work sets the stage for a systematic examination of beyond DMFT correlation effects ($d$-wave superconductivity, magnetism) on the Nd bands.

\begin{acknowledgements}
H.L and A.S.B acknowledge the support from NSF Grant No. DMR 2045826. The Flatiron Institute is a division of the Simons Foundation.
\end{acknowledgements}

\bibliography{ref.bib}
\end{document}